\documentstyle[psfig,twocolumn]{cupconf}

\ifoldfss
\else
  \ifnfssone
    \newmathalphabet{\mathit}
      \addtoversion{normal}{\mathit}{cmr}{m}{it}
      \addtoversion{bold}{\mathit}{cmr}{bx}{it}
    \newmathalphabet{\mathcal}
      \addtoversion{normal}{\mathcal}{cmsy}{m}{n}
    \else
    \ifnfsstwo
    \fi
  \fi
\fi

\def\hexnumber#1{\ifcase#1 0\or1\or2\or3\or4\or5\or6\or7\or8\or9\or
 A\or B\or C\or D\or E\or F\fi }

\makeatletter
\ifx\CUP@mtlplain@loaded\undefined
\else
\fi
\makeatother
\makeatletter
\ifx\CUP@mtlplain@loaded\undefined
\else
\fi
\makeatother
\title[JENAM 2000 Proceedings]{Cosmological Thermal Decoupling and 
Primordial Molecules\footnote{Talk given at the JENAM, May 29th-June 3rd 2000, 
Moscow (Russia)}}

\author[D. Puy]
{Denis Puy}

\affiliation{Institute of Theoretical Physics, University of Zurich, 
8057 Zurich (Switzerland)
\\
and 
\\
Paul Scherrer Institute, Laboratory for Astrophysics, 
5232 Villigen (Switzerland)
\\
puy@physik.unizh.ch}

\begin{document}
\ifnfssone
\else
  \ifnfsstwo
  \else
    \ifoldfss
      \let\mathcal\cal
      \let\mathrm\rm
      \let\mathsf\sf
    \fi
  \fi
\fi

\maketitle

\begin{abstract}
Primordial chemistry began, at the recombination epoch, when the adiabatic 
expansion caused the temperature of the radiation to fall below 4000K. The 
chemistry of the early Universe involves the elements hydrogen, its isotope 
deuterium, helium with its isotopic forms and lithium. In this contribution 
I will discuss the influence of the primordial molecules on the cosmological 
decoupling. 
\\
In the framework of the gravitational instability theory, each protostructure 
started as a tiny local overdensity. As long as these inhomogeneities 
are small, their evolution can be studied by the classical linear perturbation 
theory. Once the deviations become large, the linear theory is no more valid. 
we present the role played by these molecules on the transition between the 
linear regime and the non-linear regime, and show that the molecules can 
lead to a thermal change at the turn-around point between these two regimes. 
\end{abstract}

\firstsection 
\section{Introduction}
At early times the Universe was filled up with an extremely dense and hot gas. 
Due to the expansion it cooled below the binding energies of hydrogen, 
deuterium, helium, lithium which led to the formation of 
these nuclei (Sarkar 1996). After this nucleosynthesis period the recombination process 
is not instantaneous because the electrons, captured into different atomic 
energy levels, could not cascade instantaneously down to the ground state. 
Atoms reached the ground state either through the 
cosmological redshifting of the Lyman $\alpha$ line photons or by the 
$2s-1s$ two photons process. Nevertheless the Universe expanded and cooled 
faster than recombination could completed, and small fraction of free 
electrons and protons remained.
\\
The principles of calculations of the primordial recombination have been 
mentionned initially by Shklovskii (1967), Novikov \& Zel'dovich (1967). 
Peebles (1968) was the first to 
present a theory in which the very complicated recombination process is 
reduced to simpler terms (see Puy \& Signore 2000 for the historical 
description and references therein). 
\\
At the end of the recombination 
period, it is plausible to imagine that molecules could be formed. 
Temperature of matter and radiation as well as the density are not so high and 
we have possible collisional reactions between the species. However, in 
this cosmological context we have metal-free 
gas which does not allow the efficient reaction of adsorption on the surface 
of grains. 
\\
The literature on the chemistry in the post-recombination epoch has 
grown considerably in the recent years.  Many authors have developped 
studies of primordial chemistry in different contexts. For example Lepp \& 
Shull (1984), Latter \& Black (1991), Puy et al. (1993), Stancil et al. 
(1996), Galli \& Palla (1998) for the chemical network; 
Palla, Galli \& Silk (1995), Puy \& Signore 
(1996, 1997, 1998a, 1998b), Abel et al. (1997) and Galli \& Palla (1998) in 
the context of the formation of the first objects. 
\\
Chemistry of the early 
Universe is the gaseous 
chemistry of the hydrogen, helium, lithium and electrons species. The 
efficiencies of the molecular formation processes is controlled by collisions, 
matter temperature and temperature of the cosmic microwave 
background radiation. The complete chemical network consists of 90 
reactions (see Puy et al. 1993 and more recently Galli \& Palla 1998 and 
references therein).
\\
Molecules could lead to radiative processes through the excitation of the 
rotational levels, and contribute to the thermal evolution of the medium. Puy et 
al. (1993) examined the thermal balance in the early Universe and showed 
that the molecular 
cooling and heating functions have a slight influence on the decrease of the 
temperature of the matter. Nevertheless molecules could play a more decisive 
role in the collapse of protoclouds, since molecular cooling can trigger 
thermal instability prior the gravitational instability, as mentionned Puy \& 
Signore (1996, 1998a, 1998b). Protocloud evolute in three phases:
\begin{itemize}
\item linear evolution which approximatively follows expansion.
\item {\it turn-around} epoch when the protocloud reaches its maximum value.
\item non-linear evolution of the collapse.
\end{itemize}
At the turn-around point it is crucial to know the initial condition and 
particularly the temperature of the pre-collapse. In this contribution, 
I will recall the thermal function generated by the primordial molecules 
(Sect. 2), then I will show the influence on the thermal decoupling (Sect. 3). In 
Sect. 4, the role of the molecules on the turn-around point will be 
discussed and a brief outlook will be presented in Sect. 5.
\section{Molecular thermal function}
The disappearance of free charged particles reduces the scattering 
cross-section (Thomson scattering). The photons decouple from the rest of the 
matter; this happens at $T=T_{dec}\sim 0.26$ eV where $T_{dec}$ is the 
temperature of the matter corresponding to the redshift $z_{dec}\sim 1100$. 
After the recombination, although the density-decrease acts against molecular 
formation, it turns out that the temperature is small enough for this 
formation occur.
\\
The chemical composition of the primordial gas consists of electrons, protons, 
hydrogen ($H$, $H^-$, $H_3^{+}$, $H_2^+$, $H_2$), deuterium ($D$, $D^+$, $HD$, 
$HD^+$, $H_2D^+$), helium ($H_e$, $He^+$, $He^{++}$, $HeH^+$) and lithium 
($Li$, $Li^+$, $Li^-$, $LiH$, $LiH^+$). Thus from the abundances of
 primordial atoms, given by the standard model of nucleosynthesis 
(see Table \ref{tab:nucleo}), we integrate the network of coupled chemical 
equations which is an initial value problem for stiff differential equations. 
Lepp \& Shull (1984), Puy et al. (1993) and more recently Galli \& Palla 
(1998) calculated, as function of redshift, the fractional abundances 
starting at the redshift $z\sim 10^4$ where $He$, $H$, $D$ and $Li$ are 
fully ionized. 

\begin{table*}[h!tbp]
\renewcommand{\arraystretch}{1.0}
\centering
\begin{tabular}{||c|c|c||} \hline Helium: $He/H$ & Deuterium: $D/H$ & Lithium 
$Li/H$ \\ \hline
$\sim 8 \times 10^{-2}$ & $\sim 4.3 \times 10^{-5}$ & 
$\sim 2.4 \times 10^{-10}$ \\ \hline
\end{tabular}
\caption{Initial abundances given by the standard of the nucleosynthesis (see 
Sarkar 1996).}
    \label{tab:nucleo}
\end{table*}

The reaction rates depend on the temperature, thus the temperature and density 
evolution equation must be solved simultaneously, which needs in turn 
the simultaneous determination of molecular 
cooling and heating rates (see Puy et al. 1993, Galli \& Palla 1998). 
The abundances of the main molecules are obtained in Table \ref{tab:molec}.

\begin{table*}[h!tbp]
\renewcommand{\arraystretch}{1.0}
\centering
\begin{tabular}{||c|c|c|c||} \hline 
$e^-/H$ & $H_2/H$ & $HD/H$ & $LiH/H$\\ \hline
$\sim 3 \times 10^{-4}$ & $\sim 10^{-6}$ & $1.2 \times 10^{-9}$ & 
$\sim 7 \times 10^{-20}$ \\ \hline
\end{tabular}
\caption{abundances of primordial molecules at $z =5$.}
    \label{tab:molec}
\end{table*}

The finite amount of primordial molecules, such as $H_2$, $HD$ and $LiH$ 
formed immediately after the recombination of cosmological hydrogen 
can induce a thermal response 
of molecules on the medium through the interaction with the radiation. 
The presence of non-zero permanent electric dipole moment makes $HD$ and 
$LiH$ a potentially more important coolant than $H_2$ at modest temperatures, 
although $HD$ and $LiH$ are much less abundant than $H_2$. Below 3000 K, 
only the rotational levels of the molecules can be excited. The 
population of the rotational levels is mainly due to collisional excitation 
and de-excitation with $H$, $H_2$ and $He$ on one hand and to 
radiative processes (absorption from cosmic background radiation and 
spontaneous or induced emission) on the other hand. Molecular cooling 
corresponds to collisional excitation followed by radiative transition, 
provided that no further absorption occurs. Molecular heating is due to 
the radiative excitation from cosmic microwave background followed by 
collisional de-excitation. Notice that, although the radiative de-excitation 
is faster than collisional one (because the excitation have the opposite 
ordering), the full cooling and heating processes must be evaluated. 
Molecular thermal function $\Psi_{molec}$, characterized by the two 
processes (molecular heating and cooling), is defined by:

\begin{equation}
\Psi_{molec} \, = \, \sum_{k} \, 
\Bigl[ \Psi_k \Bigr]
\end{equation}   

\noindent
where the index $k$ is defined for each molecule ($H_2$, $HD$ or 
$LiH$) and $\Psi_k$ is the molecular thermal function for the molecule $k$ 
defined by:
\begin{equation}
\Psi_k \, = \, \sum_j n_j \sum_i n_i \Bigl(
B_{ij} u^{ij} P^c_{ij} 
-
n_x C^x_{ij} P^r_{ij} \Bigr) \epsilon_{ij}
\ \ {\rm in \ erg \ cm^{-3} \ s^{-1}}
\label{eq:func}
\end{equation} 
where $n_j$ and $n_i$ is respectively the population of the rotational level 
$j$ and $i$. $C^x_{ij}$ is the rate of collision with the species $x$ 
(with the density $n_x$), $B_{ij}$ the second 
Einstein coefficient, $u^{ij}$ the radiative density of cosmic microwave 
background radiation at the energy 
$\epsilon_{ij}$, which corresponds to the transition between the levels 
$i$ and $j$; $P^c_{ij}$ and $P^r_{ij}$ define respectively the probability 
of collisional de-excitation and the probability of radiative de-excitation 
(here we consider 10 rotational levels for each molecule). The first 
term  of Eq. (\ref{eq:func}) corresponds to molecular heating when the second 
term corresponds to molecular cooling.

\section{Cosmological thermal decoupling}
\begin{figure}[h]
\centerline{\psfig{figure=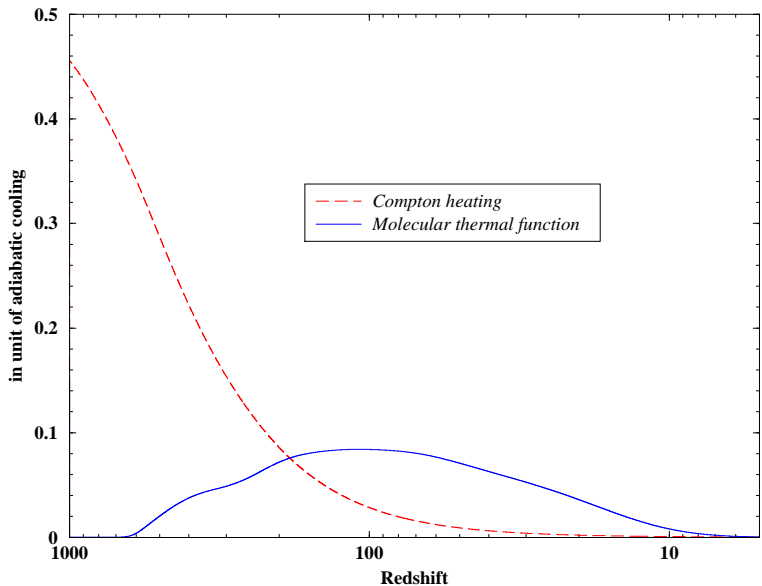,width=10cm,angle=-90}}   
\caption{Evolution of the thermal function, i.e. Compton heating 
$\Gamma_{compt}$ and molecular $\Psi_{molec}$ in unit of adiabatic 
cooling $\Lambda_{ad}$.}
\label{fig:cooling}
\end{figure}

The evolution of the energy density $u_{gas}$ of a 
homogeneous gas is described by the equation:

\begin{equation}
du_{gas} \, = \, d \Bigl( \frac{3}{2} n k T_m \Bigr) \, = \, {\cal Q} dt
\label{eq:gas}
\end{equation}
\noindent
where $k$ is the Boltzmann constant, $n$ the matter density, 
$T_m$ the temperature of matter and 
${\cal Q}$ the external energy density which depends 
on the molecular thermal function $\Psi_{molec}$, and on the net transfer of 
energy $\Gamma_{compt}$ from the cosmic background radiation to the gas via 
Compton scattering of cosmic background photons on electrons. 
Thus we have :
\begin{equation}
{\cal Q} \, = \, \Psi_{molec} +\Gamma_{compt},  
\end{equation}
where (see Peebles 1968):
\begin{equation}
\Gamma_{compt} \, = \, \frac{4 \sigma_T a_{bb} T_r^4}{m_e c} n_e 
\bigl( T_r -T_m).
\end{equation}
$\sigma_T$ is the Thomson cross-section, $a_{bb}$ is the black body constant, 
$c$ the speed light, $m_e$ the mass of electron and $n_e$ the electronic 
density. $T_r$ is the temperature of the 
radiation. Thus Eq. (\ref{eq:gas}) leads to the 
evolution of matter temperature $T_m$:
\begin{equation}
\frac{3}{2}nk \, \frac{dT_m}{dt} \, = \, -\Lambda_{ad} + \Gamma_{compt} + 
\Psi_{molec}, 
\end{equation}
where $\Lambda_{ad}$ defines the adiabatic cooling due to the expansion of the 
Universe:
\begin{equation}
\Lambda_{ad} \, = \, 3 n k T_m \, H_o (1 +z )^{3/2},
\end{equation}
$H_o$ is the Hubble constant (hereafter $H_o = 67$ Km s$^{-1}$ 
Mpc$^{-1}$) and $z$ the redshift. We have neglected the chemical heating and 
cooling due to the enthalpy of reactions (see Puy et al. 1993).
\\
We adopt the standard cosmological model with zero cosmological constant 
(Einstein-De Sitter Universe). Thus the matter temperature $T_m$ and the 
radiation temperature $T_r$ remained the same until a redshift of about 
1100, the beginning of the recombination era, after which Compton 
scattering was no longer able to overcome the cooling by expansion and $T_m$ 
fell below $T_r$. In this context the adiabatic cooling becomes dominant.
\\
In Fig \ref{fig:cooling} Compton heating 
and thermal molecular function (in unit of adiabatic cooling) are compared. 
The thermal molecular function, which is mainly a molecular heating, dominates 
the Compton heating when $z<180$. This fact should be important for the 
evolution of the matter after this redshift, particularly for the collapse 
of the first objects.

\section{Turn-around period and proto-collapse} 
\begin{figure}[h]
\centerline{\psfig{figure=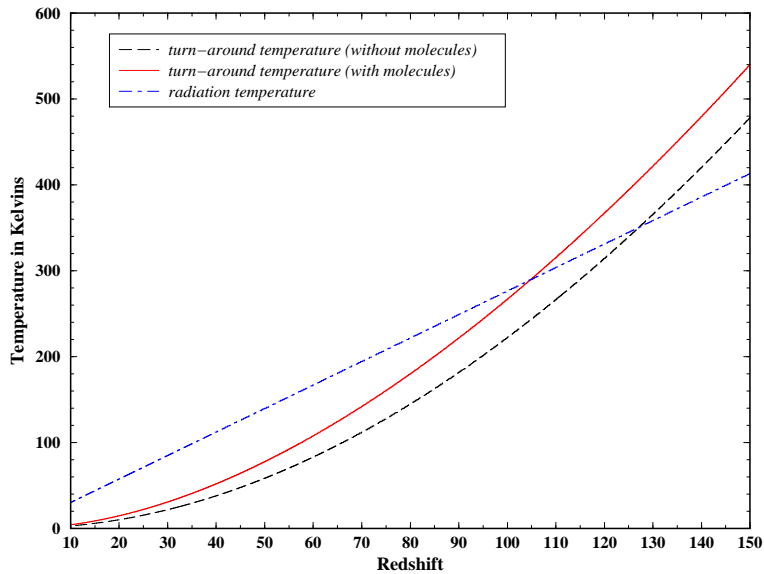,width=10cm,angle=-90}}   
\caption{Evolution of the turn-around temperature (with and without 
molecules) and of the radiation temperature.}
\label{fig:temp}
\end{figure}

We assume that, at some time in the past, there 
were small deviations from homogeneity in our Universe. These deviations 
can grow due to gravitational instability. As long as these inhomogeneities 
are small, their evolution can be studied by the classical linear perturbation 
theory. Once the deviations from the homogeneous Universe become large, the 
linear theory is no more valid. It is reasonable to expect that regions which 
are significantly overdense will collapse and eventually form gravitationally 
bound objects. In these overdense regions, the self-gravity of the local mass 
concentration will work against the expansion of the Universe; i.e. this 
region expands at a progressively slower rate compared to the background 
Universe. Such a slowing down will increase the density contrast between the 
overdense region and the background Universe and, consequently, make the 
gravitational potential of the local mass concentration more and more 
dominant. Eventually, such a region will collapse under its own self-gravity 
and will form a bound system. The details of the above process will depend 
on the initial density profile. The simplest model which one can study 
analytically is based on the assumption that the overdense region is 
spherically symmetric. 
\\
Peebles (1980) described the {\it transition} 
between the linear regime (where the expansion of the perturbation is 
maximum) and the non-linear regime (where the perturbation begins to 
collapse) by introducing the turn-around point. 
\\
In an Einstein-De Sitter Universe, the temperature at the turn-around point 
(hereafter turn-around temperature) is given by 
\begin{equation}
T_{turn} \sim \Bigl( \frac{3 \pi}{4} \Bigr)^{4/3} \, T_{m}(z_{ta}) 
\end{equation}
where $T_{m}(z_{ta})$ is the temperature of the matter at the redshift of the 
turn-around $z_{ta}$. Generally it is assumed an isothermal perturbation 
described by the initial mass spectrum:
\begin{equation}
\frac{\delta \rho}{\rho} \, = \, \Bigl[ \frac{M}{M_\star} \Bigr]^\alpha 
\, (1+z)^{-1}, 
\end{equation}
where $M$ is the mass of an overdense region (i.e. the mass of a cloud) and 
$M_\star = 10^{15}$ M$_\odot$ is a typical mass of a supercluster. Gott \& 
Rees (1975) take $\alpha \sim -1/3$. Thus the redshift of the turn-around 
is given by:
\begin{equation}
z_{ta} \, = \, \Bigl( \frac{3 \pi}{4} \Bigr)^{-2/3} \, 
\Big[\frac{M}{M_\star}\Bigr]^{\alpha}.
\end{equation}
\noindent
It is crucial to know the temperature of the matter at the turn-around 
redshift in order to evaluate the thermal initial condition of the collapse. 
We consider the same order of magnitude for the mass of the fluctuations 
given by Lahav (1986) and Puy \& Signore (1996) which correspond to 
turn-around redshift in the range: $10 < z_{ta} <150$, which is the 
typical period of the formation of the first objects. 
\\
The evolution of turn-around temperature 
is shown in Fig. \ref{fig:temp} without and with the molecular 
contributions, the evolution of 
radiation temperature is also plotted. We see two regimes 
for the two turn-around temperatures: one corresponds to a temperature higher 
than the radiation temperature $T_r$, the second is relative to temperature 
below $T_r$. It is important to estimate the redshift 
$z_i$ where the turn-around temperature is equal to $T_r$, this 
point govern the initial behaviour of the gas: 
if the turn-around temperature is below the 
radiation temperature the gas heats, when the gas cools when the 
radiative temperature is below the turn-around temperature. Thus this 
condition is located at the redshift $z_i \sim  130$  in the case without 
molecules and $z_i \sim 105$ in the case with molecules. 
Another interesting point is the point which corresponds to the turn-around 
redshift $z = 55$. This redshift corresponds to a typical value 
10$^9$ M$_\odot$ of a collapsing mass (see Puy \& Signore 1997). We find 
(see Fig \ref{fig:temp}) that the turn-around temperature $T_{turn} \sim 70$ K 
without molecules and $T_{turn} \sim 100$ K for the turn-around temperature 
with molecules. The consequences could be important if we notice that the 
lower excitation temperature for $HD$ (main thermal molecular function) 
is 112 K (close to 100 K) .

\section{Outlook}
This work shows a possible thermal modification, due to the primordial 
molecules, at the beginning of the collapse. This thermal mechanism could 
modify the initial conditions of a gravitational collapse. The 
following analysis could reveal processes of fragmentation trigerred by the 
thermal molecular function (see Puy \& Signore 1997, 1998b). This last point 
is crucial for the estimation of the amplitude of secondary cosmic 
microwave background anisotropies. Dubrovich (1977, 1993) showed that resonant 
elastic scattering must be considered as the most efficient process in 
coupling matter and radiation at high redshift. He noted that the cross 
section for resonant scattering between the cosmic microwave background and 
molecules is several orders of magnitude larger than Thomson scattering, 
even with a modest abundance of primordial molecules. 
Let us only emphasize that, again at present times, the primordial molecules abundance and the reaction 
rate of the chemical reaction are both quite uncertain.

\begin{acknowledgments}
The author gratefully acknowledge the JENAM organizers for organizing such a 
pleasant conference. I would like to thank Andrei Doroshkevich and Victor 
Dubrovich for valuable and fruitful comments and discussions. Part 
of this work has been supported by the {\it Dr Tomalla Foundation} and by the 
Swiss National Science Foundation.
\end{acknowledgments}


\begin{thebibliography}{} 

  \bibitem[]{} 
     {\sc Abel T., Anninos P., Zhang Y., Norman M.,} 1997 
     {\it New Astr.} {\bf 2}, 181. 

 \bibitem[]{} 
     {\sc Dubrovich V.,} 1977 
     {\it Sov. Astr. Lett.} {\bf 3}, 128. 

 \bibitem[]{} 
     {\sc Dubrovich V.,} 1993 
     {\it Astron. Lett.} {\bf 19}, 53. 

 \bibitem[]{} 
     {\sc Gott III R., Rees M.,} 1975 
     {\it A\& A} {\bf 45}, 265.

 \bibitem[]{} 
     {\sc Galli D., Palla F.,} 1998 {\it A \& A} {\bf 335}, 403

 \bibitem[]{} 
     {\sc Lahav O.,} 1986 
     {\it MNRAS} {\bf 220}, 259.

 \bibitem[]{} 
     {\sc Latter W., Black J.,} 1991 
     {\it ApJ} {\bf 371}, 161.

  \bibitem[]{} 
     {\sc Lepp S., Shull M.,} 1984 
     {\it ApJ.} {\bf 280}, 465. 

 \bibitem[]{} 
     {\sc Novikov I., Zel'dovich Ya.,} 1967 
     {\it Ann. Rev. Astr. Ap.} {\bf 5}, 627. 

 \bibitem[]{} 
     {\sc Palla F., Galli D., Silk J.,} 1995
     {\it ApJ} {\bf 451}, 44.             

 \bibitem[]{} 
     {\sc Peebles P.J.E.,} 1968 
     {\it ApJ} {\bf 153}, 1. 

 \bibitem[]{} 
     {\sc Peebles P.J.E.,} 1980 
     in: {\it The large scale structure of the Universe} (Princeton 
University Press)

 \bibitem[]{} 
     {\sc Puy D., Alecian G., Lebourlot J., Leorat J., Pineau des Forets,} 1993
     {\it A\& A } {\bf 267}, 337.

 \bibitem[]{} 
     {\sc Puy D., Signore M.,} 1996
     {\it A\& A } {\bf 305}, 371.

 \bibitem[]{} 
     {\sc Puy D., Signore M.,} 1997
     {\it New Astr.,} {\bf 27}, 622.

 \bibitem[]{} 
     {\sc Puy D., Signore M.,} 1998a
     {\it New Astr.,} {\bf 3 }, 27.

 \bibitem[]{} 
     {\sc Puy D., Signore M.,} 1998b
     {\it New Astr.,} {\bf 3 }, 247.

\bibitem[]{}
    {\sc Puy D., Signore M.,} 2000
    {\it New Astr. Rev.,} accepted 

  \bibitem[]{} 
     {\sc Sarkar S.,} 1996  {\it Rep. Prog. Phys-} {\bf 59}, 1493. 

 \bibitem[]{} 
     {\sc Shklovskii I.,} 1967 
     in Proceedings {\it 4th Texas Conf. on Relativistic Ap.} {\bf 5}, 627. 

 \bibitem[]{} 
     {\sc Stancil P., Lepp S., Dalgarno A.,} 1996 
     {\it ApJ} {\bf 458}, 401.

\end{thebibliography}
\end{document}